# Energy-Efficient and Robust Associative Computing with Electrically Coupled Dual Pillar Spin-Torque Oscillators

Mrigank Sharad, Deliang Fan, Karthik Yogendra and Kaushik Roy

School of Electrical and Computer Engineering, Purdue University, West Lafayette, Indiana

47907, USA

Dynamics of coupled spin-torque oscillators can be exploited for non-Boolean information processing. However, the feasibility of coupling large number of STOs with energy-efficiency and sufficient robustness towards parameter-variation and thermal-noise, may be critical for such computing applications. In this work, the impacts of parameter-variation and thermal-noise on two different coupling mechanisms for STOs, namely, magnetic-coupling and electrical-coupling are analyzed. Magnetic coupling is simulated using dipolar-field interactions. For electrical-coupling we employed global RF-injection. In this method, multiple STOs are phase-locked to a common RF-signal that is injected into the STOs along with the DC bias. Results for variation and noise analysis indicate that electrical-coupling can be significantly more robust as compared to magnetic-coupling. For room-temperature simulations, appreciable phase-lock was retained among tens of electrically coupled STOs for up to 20% 3σ random variations in critical device parameters. The magnetic-coupling technique however failed to retain locking beyond ~3%  3σ parameter-variations, even for small-size STO clusters with near-neighborhood connectivity. We propose and analyze Dual-Pillar STO (DP-STO) for low-power computing using the proposed electrical coupling method. We observed that DP-STO can better exploit the electrical-coupling technique due to separation between the biasing RF signal and its own RF output.

## I. Introduction

A Spin-Torque Oscillator has two ferromagnetic layers separated by either a thin non-magnetic metal (Giant Magneto Resistance-GMR device) or a thin insulating oxide (Tunneling Magneto Resistance-TMR device) layer (Fig. 1a). The ferromagnetic layers have two stable spin-polarization states, depending upon magnetic anisotropy [1]. The magnetization of one of the layers is fixed, while that of the other (free-layer) can be influenced by a charge current passing through the device or by an applied magnetic field. The high-polarity fixed magnetic-layer spin-polarizes the electrons constituting the charge-current, which in turn exert spin transfer torque (STT) on the free-layer [2]. A static magnetic field can be used to obtain sustained spin-precession of the free layer at an angle φ, at which the STT and the damping torque balance out each other. (Fig 1(a)) [3]. The resistance of the spin valve can be expressed as a function of relative angle between the spin-polarization of the two ferromagnetic layers θ as

$$R = \left(\frac{R_P + R_{AP}}{2}\right) + \left(\frac{R_P - R_{AP}}{2}\right)\cos\theta \qquad (1)$$

Where $R_P$ and $R_{AP}$ denote the resistance when the two layers are parallel (θ = 0) and antiparallel (θ = 180). The absolute resistance of a GMR device is much smaller than that of a TMR device (less than ~1 ohm). A GMR based Spin-Torque Nano Oscillator (STO), being fully metallic, can be operated with very low voltage (<10 mV). However, the sensed signal amplitude is very low which requires complex sensing circuitry to amplify the signal, leading to high power consumption [4]. On the other hand, though the TMR based STO can provide large amplitude output signals, due to the high-resistance tunnel junction, it requires a large bias voltage, leading to energy inefficiency at the device level.

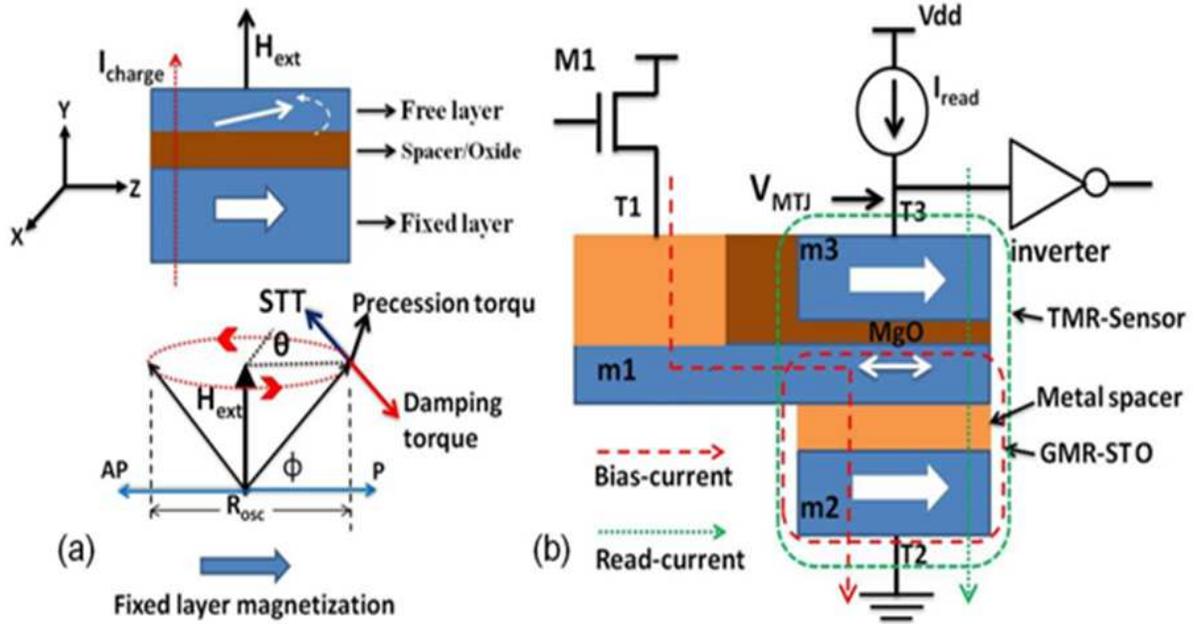

Fig 1 (a) STO operation in presense of spin-torque and magnetic field perpendicular to the plane of polarization, (b) Dual Pillar Spin Torque Nano Oscillator (DP-STO) with biasing and sensing circuit: $m_1$ is the free layer with dimensions: $30 \times 30 \times 2$ nm$^3$, $m_3$ forms the reference layer for the MTJ with area $15 \times 30$ nm$^2$.

We proposed a dual-pillar spin torque oscillator that can overcome the aforementioned bottleneck and can be suitable for energy-efficient computing [5]. Fig 1(b) shows the schematic diagram of the proposed three-terminal Dual Pillar Spin Torque Nano Oscillator (DP-STO). The DP-STO structure has an extended free layer magnet-$m_1$. Towards the right it forms a GMR interface with one fixed magnet layer-$m_2$, and, a TMR interface with another fixed magnet layer-$m_3$. A simple CMOS interface circuitry for biasing the DP-STO and sensing the oscillations is also shown in Fig 1(b). Input bias current which sets the free layer in oscillation is applied between terminals $T_1$ and $T_2$ using a transistor-$M_1$ (dashed line in Fig 1(b)). Owing to the low resistance magneto-metallic GMR channel, the bias-current can be applied through transistor $M_1$ with a very low drain-to-source voltage (transistor operating in deep triode region). This current induces spin torque on the portion of free layer in contact with GMR interface and sets the magnetization of the free layer into sustained oscillations, under an applied magnetic field-$H_{ext}$, vertical to the plane of magnetization.

Oscillations of multiple STOs can be synchronized with the help of electrical or magnetic coupling [6][7][8][9]. Electrical coupling can be achieved by injection of a common RF signal into multiple DP-STOs, to which they can acquire a phase-lock [6]. Magnetic coupling may be achieved through spin-wave interaction [7] or dipolar-coupling [8] [9]. A network of coupled STOs can be applied to associative computing task [10], [11]. However, it is important to assess the impact of thermal noise and device parameter-variations upon the dynamics of coupled oscillator to evaluate the feasibility of such computing [12], [13]. In this work we compare the performance of electrical and magnetic coupling for STOs applied to associative computing. We show that electrical coupling based on injection locking can be significantly more robust as compared to magnetic coupling techniques. We also show that DP-STO can be better suited for such an electrical coupling technique, with respect to robustness and energy efficiency.

## II. Associative Computing With Coupled STOs

Associative pattern-matching operation can be achieved using arrays of coupled STOs by exploiting their input-dependent locking characteristics [10]. Fig 2(a) shows the transient plot of two coupled STOs (solid and dashed lines) lock over time. In Fig. 2(b) current through one of the STOs is kept constant at 100µA and the current through the second STO is increased from 90 µA to 120 µA. Constant current through the first STO generates a constant frequency of oscillation, whereas, the frequency of the second device increases with its input current. When the frequencies of STOs are far apart, they oscillate independently. They acquire phase and frequency-lock when their frequencies lie in 'locking-range', as depicted in Fig 2(b). The locking range can be defined as the maximum difference between the DC biases of the two STOs for which phase-lock is retained.

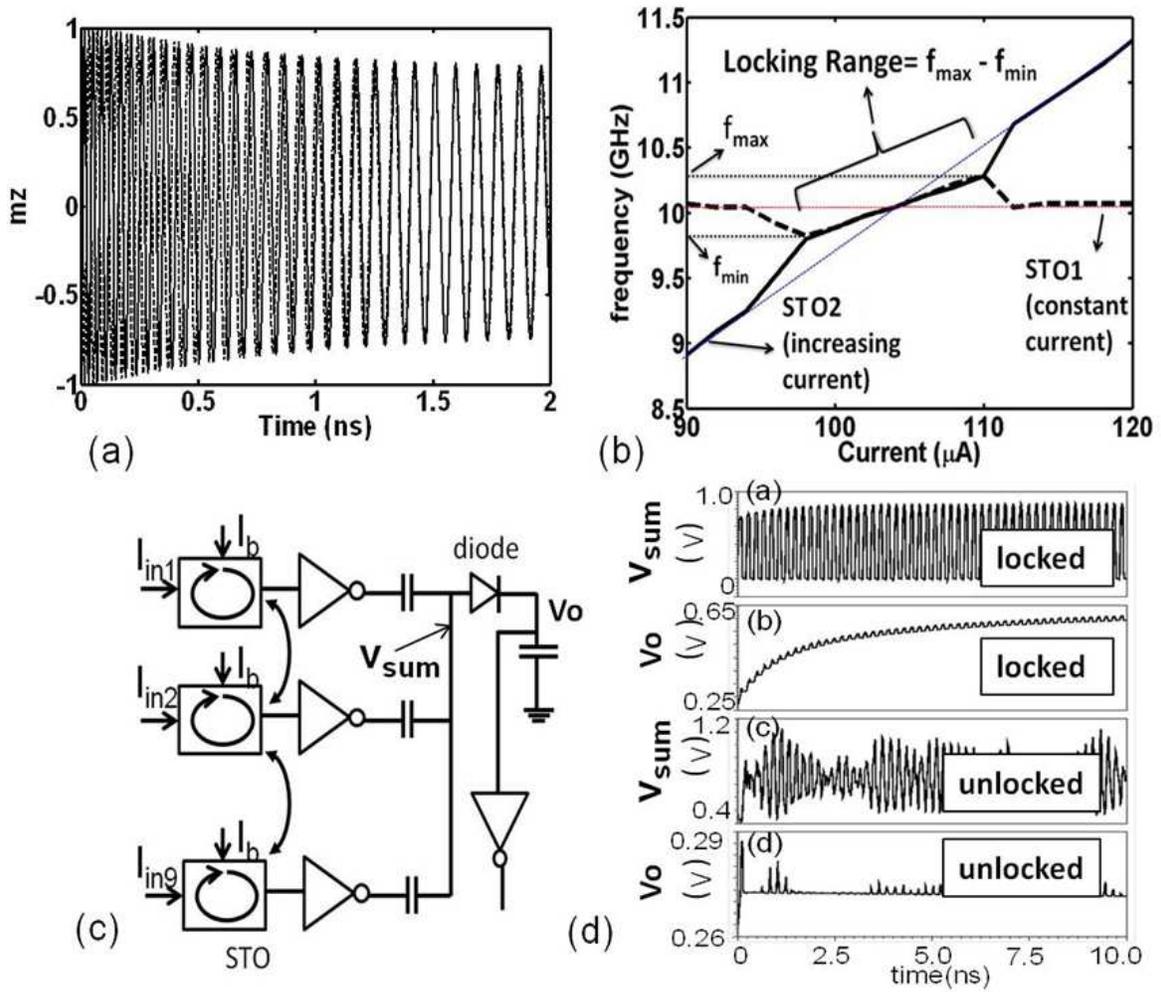

Fig. 2 (a) Transient-plot of phase-frequency locking between two STOs coupled using dipolar-interaction, (b) frequency locking range of two STOs using mono-domain simulation, matched closely with multi-domain micro-magnetic simulations (c) averager and peak-detector circuit for detecting edge-map, (d) transient response of edge-detection circuit for locked and unlocked case.

Coupled STOs can be used to evaluate the degree of match between two analog vectors. Fig. 2c shows the circuit for an STO-based associative-module (AM) that achieves this functionality. In this circuit, all the STOs are coupled (electrically or magnetically) and are biased with the same DC input. This enforces phase-locked oscillation of all the STOs in the AM. To compute the associative matching between two analog vectors of $N$ elements, current-inputs proportional to the element-wise difference of the two vectors are injected into $N$ coupled STOs. If the two vectors closely match each other, the inputs to the STOs are too small to bring them out of the locking range. The STOs therefore retain phase and frequency lock. On the other hand, if the two vectors are significantly different, the inputs to the STOs are large in magnitude resulting in loss of locking. The circuit shown in fig. 2c performs a capacitive summation of the individual STO waveforms of the AM, and applies the sum to an integrator formed by a diode-capacitor combination. In the case of phase-locked waveform, the summation results in a regular sinusoidal waveform which leads to fast charging of the integrator output (fig. 2d). On the other hand, in the case of un-locked STOs, the summation is an irregular and low amplitude waveform which leads to lower or negligible charging of the output. Thus the case of match between an input-vector and a template-vector can be identified by comparing the integrator output.

In order to simulate the matching operation for 16x16 pixel images in fig. 3a. The pixel-wise difference between the images and the stored templates were injected into the STOs in different AMs with 8-STOs each (requiring 256/8 = 32 clusters in total). The integrator outputs of all the associative modules were summed and the result was considered as the degree of match (DOM). Higher value of the integrator output implied closer match and vice versa.

Next we compare two different coupling mechanisms for STOs, namely, magnetic and electrical, for associative computing, with respect to variation and noise tolerance.

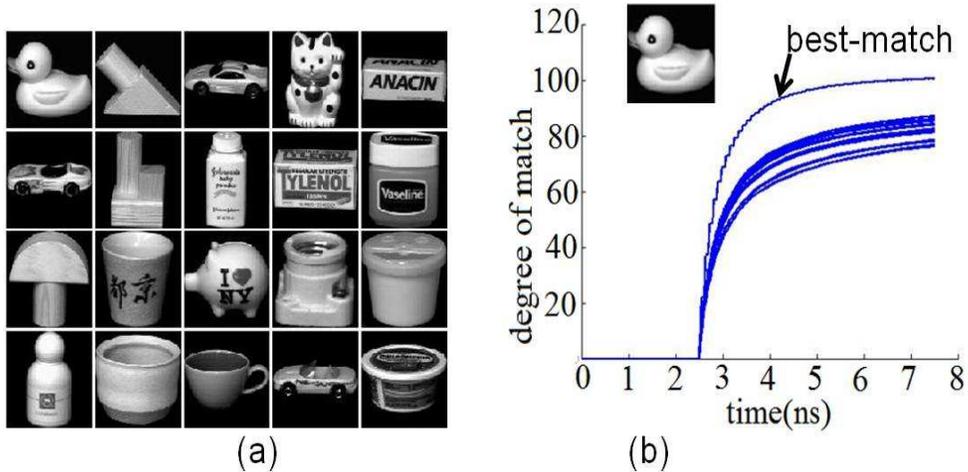

Fig. 3(a) Image-data-set used in simulation: pixel values corresponding to the individual images were stored as 1-D analog templates, (b) integrator outputs for a particular input image compared with all the other template images.

## III. Magnetic Coupling

Magnetic coupling may be achieved through spin-wave interaction [7] or dipolar-coupling [8], [9]. Spin-wave coupling may involve interaction through exchange as well as dipolar fields of oscillating magnetic domains, through a shared magnetic-substrate or channel [7]. Dipolar interaction on the other hand, can facilitate locking of physically isolated DP-STNOs lying in close proximity [8]. In this work we employ dipolar-field interaction for coupling multiple DP-STNOs.

Fig. 4a and fig. 4b show the micro-magnetic simulation plots for the locked and the un-locked cases for dipolar-field coupled STOs respectively. In fig. 4a, showing the locked case, the inputs are small and hence fail to disturb the locking due to a common DC bias and near-neighbor dipolar-filed interaction. The average magnetization for this case is shown in fig. 4c.

The inputs in the case of the unlocked oscillations, shown in fig. 4b are large enough to overcome the locking, resulting in irregular average waveform, as shown in fig. 4d.

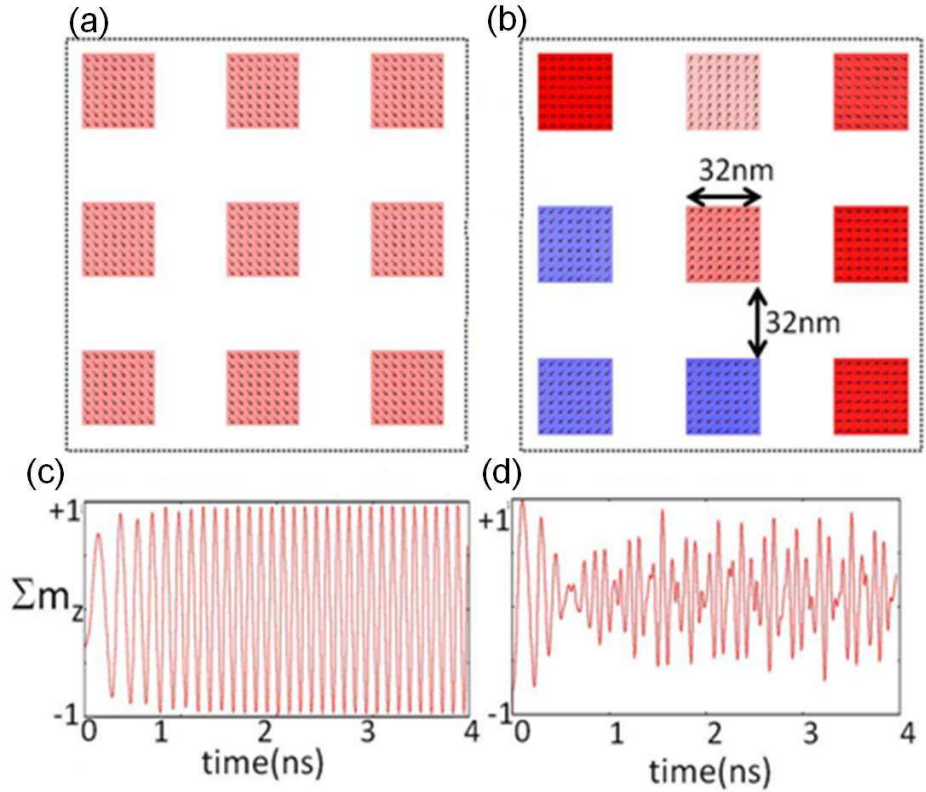

Fig. 4 Micro-magnetic simulation plots for a 3x3 STO array with dipolar coupling (a) for locked case, (b) unlocked case; evolution of average magnetization for the cluster (c) in Fig.4a, (d) in fig. 4b.

We estimated the impact of parameter variation by introducing Gaussian spread in the critical STO parameters like the saturation magnetization $M_s$ and the Gilbert damping constant $\alpha$. These parameters can have significant spread across multiple device-samples and hence it is important to evaluate the impact of spread in these parameters upon the dynamics of coupled STOs. Towards this end, we simulated associative pattern-matching circuitry based on 9-coupled STOs as described in section-II. Fig. 5 shows that there is effectively no locking for 20% spread in these parameters, for a cluster of 9 coupled STOs. The integrator outputs for the case of parameter-spread are also compared with that of the ideal case.

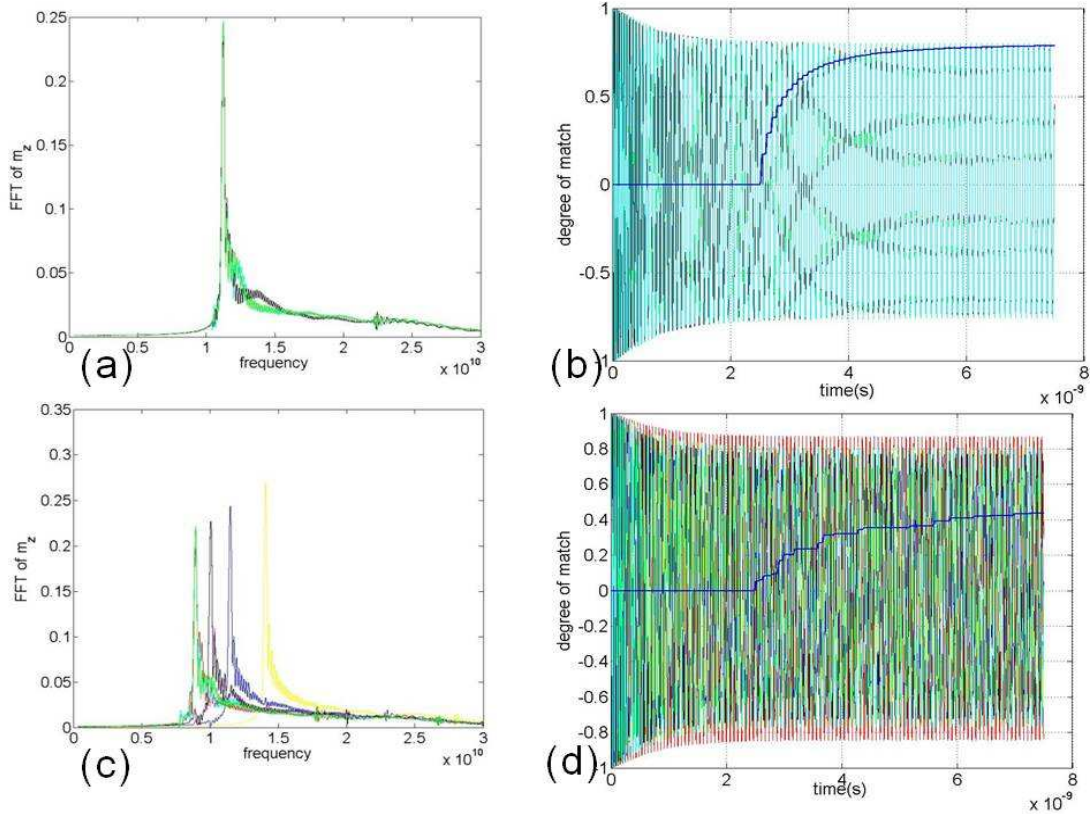

Fig. 5(a) FFT of 9 magnetically coupled STOs with identical device parameters, (b) overlapped transient waveforms with same DC bias and integrator output (deep-blue curve), (c) FFT of 9 magnetically coupled STOs with 20% spread in Ms and $\alpha$ , (d) STO waveforms and integrator output corresponding to part-c.

The associative matching operation was simulated for the image-set in fig. 3, as describe in section-II. Multiple clusters of magnetically coupled 9-STOs were used to evaluate the DOM ( which are effectively the integrator outputs of the individual clusters) for groups of 9 pixels each. The DOM of individual AMs (formed by the 9-STO clusters) were merged to get the overall

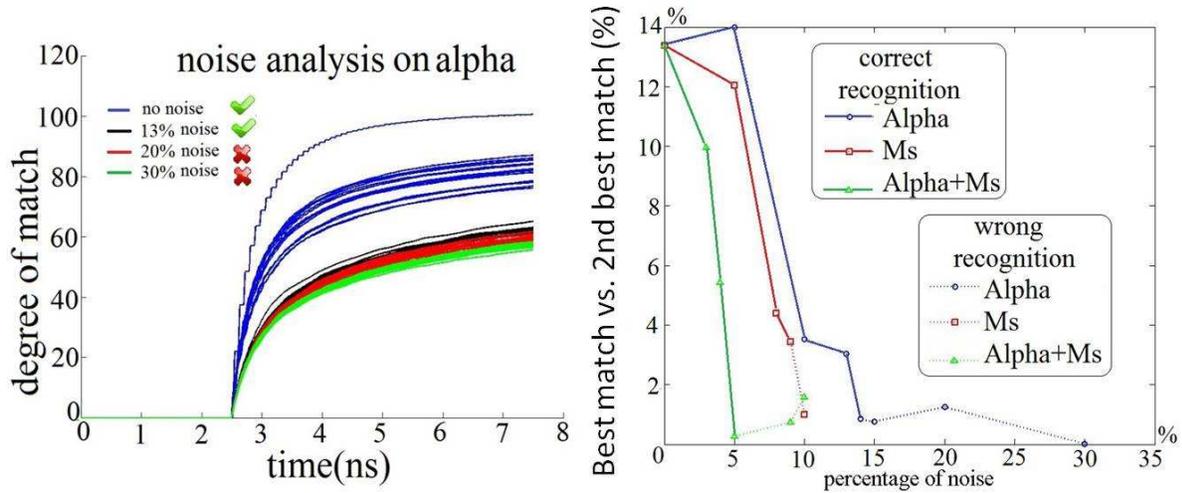

Fig. 6 (a) integrator outputs for three different degrees of parameter-spread using zero-temperature simulation, (b) % difference between the best and the second-best match of the integrator output, for increasing % variations,

DOM for the entire image. Fig. 6 a shows the effect of parameter variation on the AM outputs. It shows results for four different degrees of parameter variations. For the ideal case (with zero parameter variations), the best match-case (when the input image matches the template) is clearly distinguishable from the non-matching cases and hence can be easily detected by a coarse-comparator. With the addition of ~10% parameter variation, the best-match case was still correct (i.e, obtained the highest value), but it is too close to the rest of the outputs to be reliability detected. For further higher variations, the best-matching result was found to be incorrect. Fig. 6b shows the difference between the best and the second best matches with increasing parameter variations. The thick lines denote correct match (i.e., the best match being the correct template), whereas the thinner lines connect the points with wrong match. The plot shows that, even for zero-temperature simulations the AM based on magnetically coupled STO fails to perform correctly beyond 5% variations in α and $M_s$. Stochastic Landau-Lifshitz-Gilbert formulation was used to incorporate the effect of thermal noise in the STO-dynamics. The corresponding transient plots for AM outputs are given in fig. 7, which show that the best match case was indistinguishable beyond 2% parameter variation.

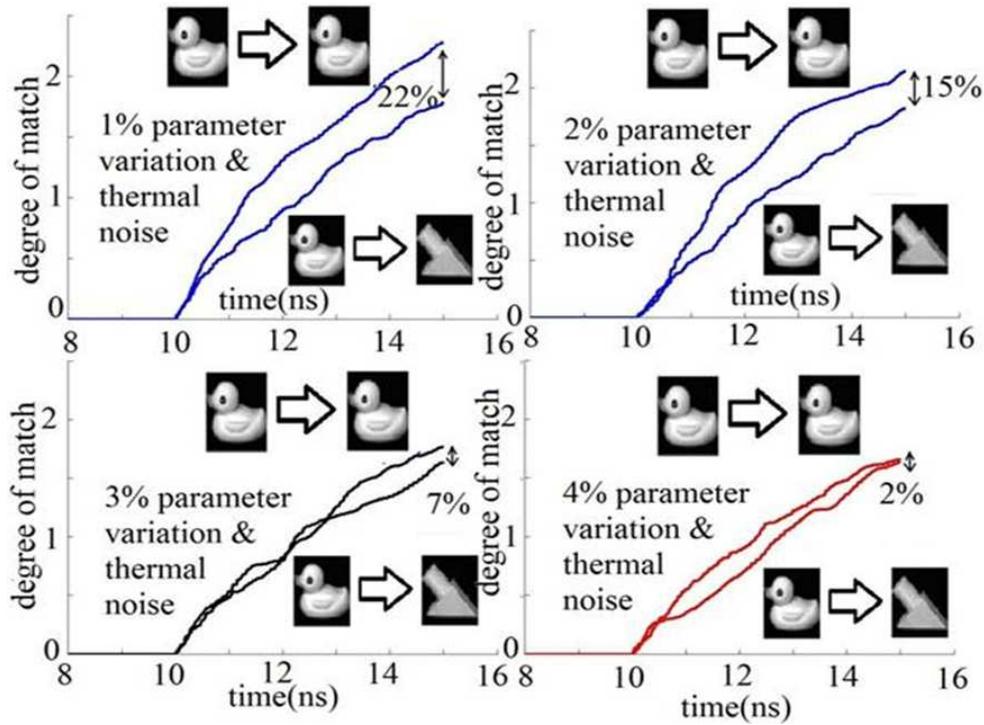

Fig. 7 Integrator waveform for best and second-best match for magnetic coupling

The foregoing analysis indicates that it might be challenging to build robust associative modules with magnetically coupled STOs due their weak immunity to thermal noise and parameter-variations. We explored an alternate coupling mechanism for STOs that can possibly offer higher robustness. This method, based on RF-injection locking is discussed next.

## IV. Electrical Coupling

In order to establish electrical locking a common RF signal can be injected into a larger number of oscillators [6]. If the RF frequency is close to that of the bias frequency of the STOs

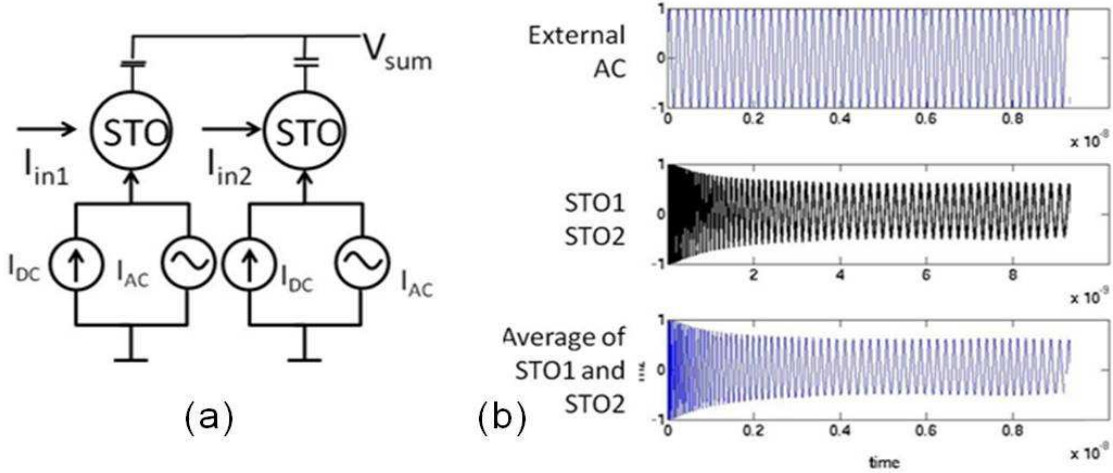

Fig. 8 (a) Two STOs with electrical coupling, (b) transient waveforms for the two STOs showing acquisition of phase-lock

Table-I DC and AC locking range with increase in AC amplitude.

| DC bias | AC-amplitude | DC-locking range | AC-locking range |
|---------|--------------|------------------|------------------|
| 50µA    | 10µA         | 5µA              | 20%              |
| 50µA    | 20µA         | 10µA             | 40%              |
| 50µA    | 40µA         | 15 µA            | 60%              |

(determined by the DC bias), they acquire phase-lock to the injected signal. Fig. 8a pictorially depicts this scheme for two STOs. In this circuit, both the STOs are biased with identical DC voltages, along with identical AC signals. The frequency of the AC signal is chosen to be close to that of the STO oscillation produced with the DC bias alone. Fig. 8b shows the transient plots for the two STOs, showing phase-locking due to the injected AC signal. For a significantly wide range of AC amplitudes of the global RF signal, the STOs were found to phase lock with it, at a constant phase-difference (same for all STOs). The phase difference among the different STOs however was close to zero under ideal conditions (zero noise and parameter variations). This implied an effective mutual synchronization and phase-locking among the STOs. The DC locking range is defined as the maximum difference between the DC inputs of the two STOs for which the phase-lock is retained. Similarly AC locking range can be defined as the maximum difference in the AC-bias amplitudes of the two STOs for which phase-lock persists. Table-I gives the DC and AC locking range for different AC-magnitudes. It shows that both the locking ranges improve with increase in the AC amplitude.

The effect of increasing AC bias on the locking of 8 electrically coupled STOs is shown in fig. 9, under 5% parameter variation. The solid-line corresponds to the reference AC signal ( normalized ). The oscillation waveforms for the 8 STOs are plotted using dotted lines.

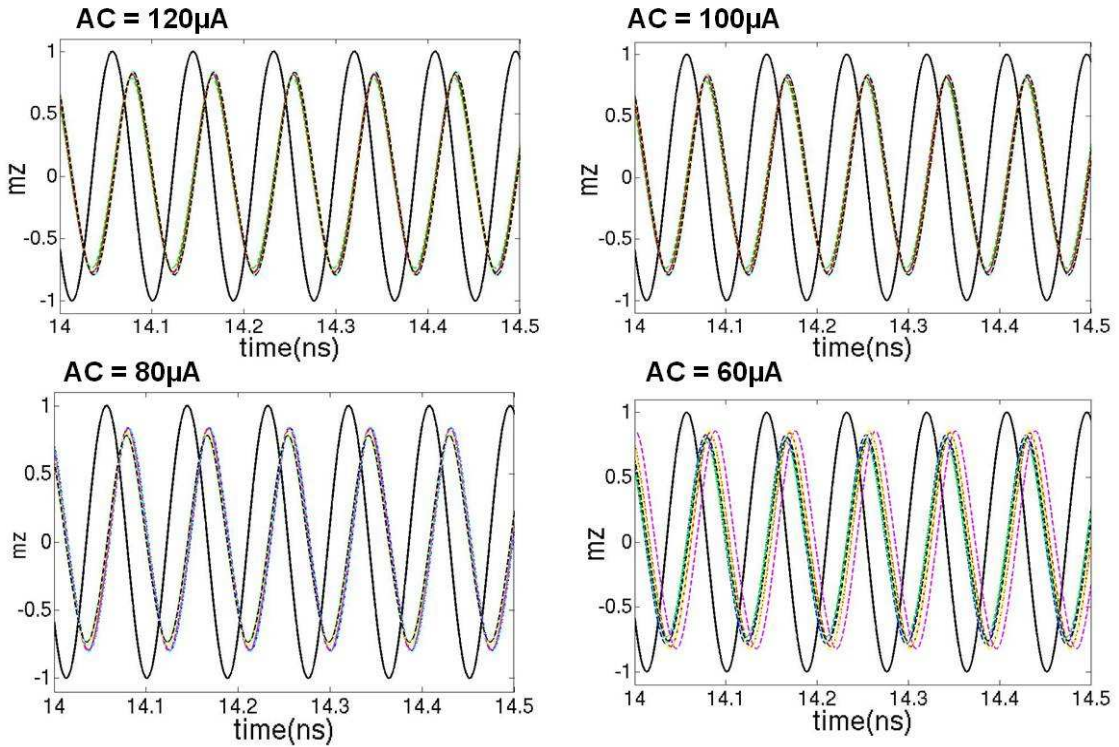

Fig. 9 Transient plots for 8 electrically coupled STOs with 5% parameter variation for different AC amplitudes.

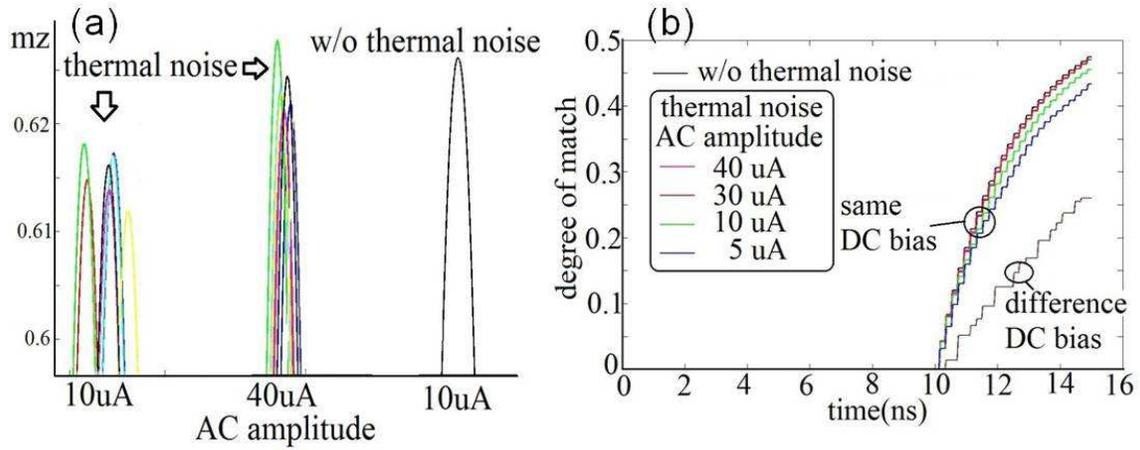

Fig. 10 (a) transient plots (zoomed in) for 8 electrically coupled STOs for two different AC amplitudes under thermal noise and 10% parameter variations, (b) increase in integrator output with increasing AC amplitude, for a given image-input.

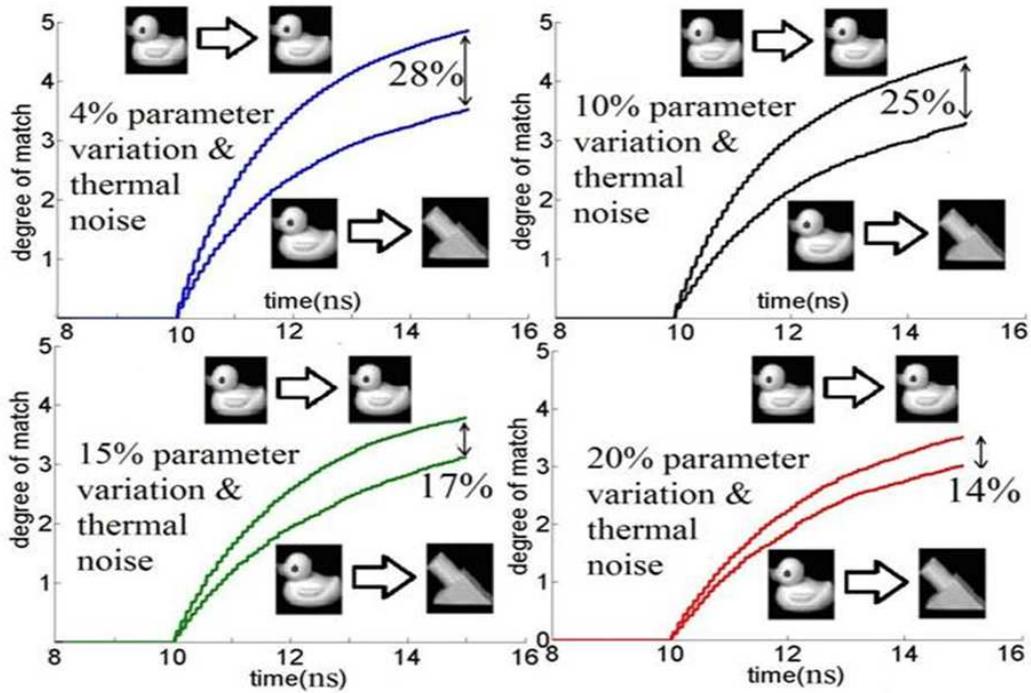

Fig. 11  Integrator waveform for best and second-best match for electrical coupling

Similar results for thermal noise are shown in fig. 10a, depicting improved phase locking for higher AC amplitude using room-temperature simulations. Fig. 10b shows the output of the integrator circuit for the electrically coupled STOs. The increase in the output value results from stronger coupling and hence cleaner averaged waveform (obtained by adding the individual STO-waveforms). Thus, stronger AC-bias leads to stronger electrical coupling and improves the tolerance to parameter variation and thermal noise.

The integrator outputs for the best and the second-best match for AM based on electrically coupled STOs are shown in fig. 11. The plots show that the associative modules could provide distinguishable outputs for up to ~20 % parameter variations, for room-temperature simulations. We used this method to couple up to 32 STOs. No significant degradation in variation tolerance was observed with increasing number of STOs.

These results indicate the superiority of the electrical coupling method over the magnetic coupling techniques. The key factor behind this advantage is the use of a common global RF signal in the case of electrical coupling, which is not influenced by the thermal noise and parameter variations of individual STOs. However, in the case of magnetic coupling, the interaction between each pair of STOs is reciprocal and hence is more prone to individual noise and parameter fluctuations. Next we discuss the benefits of DP-STO in the design of electrically coupled associative module.

## V.     DP-STO for Robust and Low Power Electrical Coupling

As described in section-II, a DP-STO employs a low-resistance GMR-interface (between free layer $m_1$ and fixed layer m2 in fig. 1b) for oscillations with low bias voltages. Low voltage biasing reduces static power consumption at the device level. A high-resistance TMR-interface (between $m_1$ and a fixed layer $m_2$) is used for sensing the free-layer oscillations using a small read current. The TMR interface can provide large-swing output signal with a small sensing current, thereby mitigating the need of additional amplification. Thus a DP-STO can achieve low power consumption along with simplified interface with CMOS. Table-II compares the power consumption of the proposed DP-STO with single-pillar GMR and TMR STO, showing more than ~98% power saving for the prior. The corresponding device parameters are given in table-III.

Table II: Comparison of Power dissipation of DP-STO with 2 terminal STO

| Device | GMR-STNO | MR-STNO | DP-STNO |
| --- | --- | --- | --- |
| Bias voltage | 10 mV | 0.7 V | 10 mV |
| STNO Power | 0.6 μW | 50 μW | 0.7 μW |
| Interface power | 1.8 mW | 0.6 μW | 0.5 μW |
| Total power | 2.3 mW | 50.6 μW | 1.2 μW |

Another important advantage offered by DP-STO is robust electrical coupling. As mentioned earlier, the proposed electrical coupling method results in a finite but constant phase difference between the global RF signal and the coupled STOs. For a 2-terminal STOs this results in a distorted output, due the mixing of the RF bias and the STO's own oscillations (which have a constant phase offset). The corresponding plots are shown in fig. 12a. As a result of this distortion the amplitude of the summed output of an STO cluster is found to be significantly lower (~50%) and has lower noise immunity. The DP-STO on the other hand provides isolated paths for the RF bias and the sensed output which is a clean sinusoid, as shown in fig. 12b. Thus, these advantages of DP-STO may be attractive for the implementation of robust and low-power associative modules.

Table III: Magnet and LLGS parameters

| Parameter | Value |
|---|---|
| Effective area of device | 40x40 nm$^2$ |
| Free layer thickness ($t_{FL}$) | 2 nm |
| Gilbert damping coefficient ($\alpha$) | 0.01 |
| Polarization factor (P) | 0.6 |
| Saturation magnetization-Ms | 800 emu/cm$^3$ |
| Energy barrier (Ea) | 10 $K_B T$ |
| Grid size | 2x2x2 nm$^3$ |

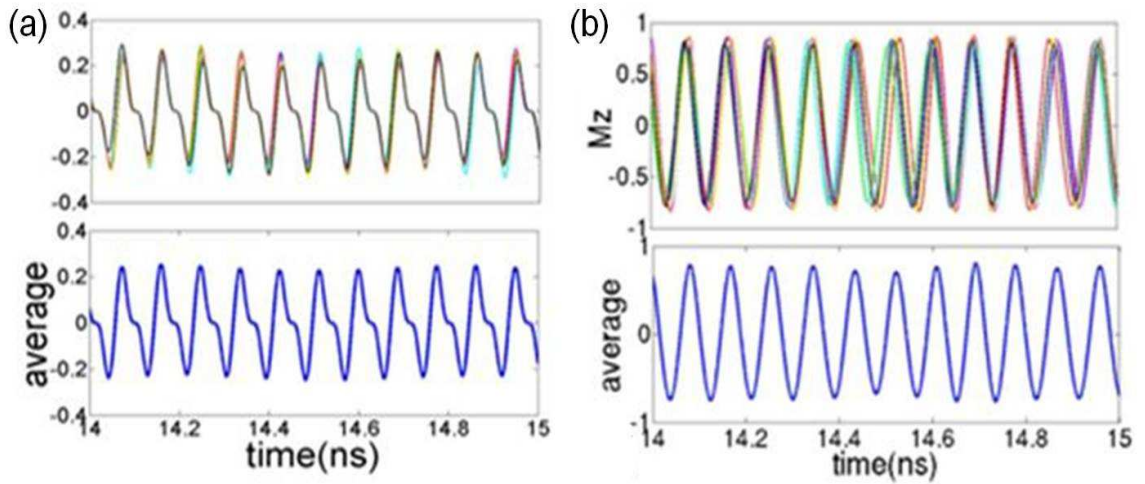

Fig. 12 (a) waveforms for electrically single-pillar coupled STOs and their average (b) waveforms for electrically coupled DP-STOs and their average.

# VI. CONCLUSION

We analyzed the impact of parameter-variation and thermal-noise on magnetic and electrical coupling mechanisms for STOs for their prospective application in non-Boolean/associative computing. Results indicate that the electrical coupling can be significantly more robust as compared to magnetic coupling techniques. We proposed and analyzed Dual-Pillar STO for low power and compact CMOS interface. We observed that DP-STO can better exploit the electrical coupling technique by due to separation between the biasing RF signal and its own RF output.

**Acknowledgement**: This research was funded in part by NSF, SRC, DARPA, MARCO, and StarNet.